\documentclass[fleqn,usenatbib]{mnras}

\usepackage{newtxtext,newtxmath}
\usepackage[T1]{fontenc}
\DeclareRobustCommand{\VAN}[3]{#2}
\let\VANthebibliography\thebibliography
\def\thebibliography{\DeclareRobustCommand{\VAN}[3]{##3}\VANthebibliography}
\usepackage{graphicx}	
\usepackage{amsmath}
\usepackage{xcolor}

\title[Classification of variable stars]{Structural properties and classification of variable stars: A study through unsupervised machine learning techniques}

\author[Paul et al.]{
Suman Paul,$^{1}$\thanks{E-mail: spappmath\_{rs}@caluniv.ac.in}
Tanuka Chattopadhyay,$^{1}$\thanks{E-mail: tchatappmath@caluniv.ac.in}\\
$^{1}$ Department of Applied Mathematics, University of Calcutta, 92 A.P.C. Road, Kolkata 700009 
}

\date{Accepted XXX. Received YYY; in original form ZZZ}

\begin{document}
\label{firstpage}
\pagerange{\pageref{firstpage}--\pageref{lastpage}}
\maketitle

\begin{abstract}
The advancement in the field of data science especially in machine learning along with vast databases of variable star projects like the Optical Gravitational Lensing Experiment (OGLE) encourages researchers to analyse as well as classify light curves of different variable stars automatically with efficiency. In the present work, we have demonstrated the relative performances of principal component analysis (PCA) and independent component analysis (ICA) applying to huge databases of OGLE variable star light curves after obtaining 1000 magnitudes between phase 0 to 1 with step length 0.001 for each light curves in identifying resonances for fundamental mode (FU) and first overtone (FO) Cepheids and in the classification of variable stars for Large Magellanic Cloud (LMC), Small Magellanic Cloud (SMC) as well as Milky Way (MW). We have seen that the performance of ICA is better for finding resonances for Cepheid variables as well as for accurately classifying large data sets of light curves than PCA. Using K-means clustering algorithm (CA) with respect to independent components (ICs), we have plotted period-luminosity diagrams and colour-magnitude diagrams separately for LMC, SMC and MW and found that ICA along with K-means CA is a very robust tool for classification as well as future prediction on the nature of light curves of variable stars.
\end{abstract}


\section{Introduction}
Over the past few decades, the importance of variable stars in the history of stellar evolution has been perceived and from the observed light curves of these variable stars, astronomers derived pieces of information not only to interpret their life histories but also that of galaxies. In the context of interpretation, analyzing the structural changes of different variable stars as well as their more accurate classification through different available statistical techniques is crucial. The available databases of variable star projects like MACHO (Massive Compact Halo Object), NSVS (Northern Sky Variability Survey), OGLE (Optical Gravitational Lensing Experiment), ASAS (All Sky Automated Survey), etc. together with satellite missions like CoRoT (Convection Rotation and Planetary Transits), Gaia, Kepler, etc. become primary resources as they contain a large amount of light curve data present in our Galaxy as well as Magellanic clouds. \\ Fourier decomposition (hereafter, FD) is a reliable, efficient and widely used technique in the study of stellar pulsation. Schaltenbrand $\&$ Tammann (\citeyear{Schaltenbrand1971}), in their paper introduced this technique for the first time to derive UBV light curve parameters of 323 Galactic Cepheids from photoelectric observations. However, Simon $\&$ Lee (\citeyear{Simon1981}) developed this method to reconstruct light curves as well as to describe the Hertzsprung progression in Cepheid light curves. The Fourier method had been applied for different purposes like investigating a large sample of light curves of RR Lyrae field stars (Simon $\&$ Teays \citeyear{Simon1982}), for analysis of the velocity curves of classical Cepheids (Simon $\&$ Teays \citeyear{Simon1983}), for the exploration of the structural properties of s-Cepheids (Antonello $\&$ Poretti \citeyear{Antonello1986}) as well as different kinds of Cepheid variables (Andreasen $\&$ Petersen \citeyear{Andreasen1987}; Andreasen \citeyear{Andreasen1988}; Simon $\&$ Kanbur \citeyear{Simon1995}; Deb $\&$ Singh \citeyear{Deb2009}), for the computation of the distance moduli of RR Lyrae stars (Kova$^\prime$cs $\&$ Jurcsik \citeyear{Kovacs1997}), for the determination of central period of Hertzsprung progression (Welch et al. \citeyear{Welch1997}; Beaulieu \citeyear{Beaulieu1998}), for a detailed study of light curve properties of RRab (fundamental mode pulsators) and RRc (first overtone pulsators) (Poretti \citeyear{Poretti2001}; Moskalik $\&$ Poretti \citeyear{Moskalik2003}), reconstructing Cepheid light curves to deal with possible metallicity dependence (Antonello et al. \citeyear{Antonello2000}; Ngeow et al. \citeyear{Ngeow2003}). Some of the recent studies demonstrated the involvement of Fourier coefficients in the determination of physical parameters like luminosity, absolute magnitude, effective temperature, metallicity, etc. as a function of period, wavelength, etc. (Deb $\&$ Singh \citeyear{Deb2010}; Nemec et al. \citeyear{Nemec2011}; Bhardwaj et al. \citeyear{Bhardwaj2015}; Bhardwaj et al. \citeyear{Bhardwaj2017}) as well as chemical and structural analysis for fundamental mode RR Lyrae stars of LMC (Deb $\&$ Singh \citeyear{Deb2014}).
Fourier decomposition technique is not perfectly adequate in the sense of enlargement of light curve databases of variable stars. However, some authors like (Hendry et al. \citeyear{Hendry1999}; Kanbur et al. \citeyear{Kanbur2002}; Kanbur $\&$ Mariani \citeyear{Kanbur2004}; Tanvir et al. \citeyear{Tanvir2005}; Sarro et al. \citeyear{Sarro2009}) applied this method as a preprocessor for principal component analysis (PCA) scheme.\\
Principal component analysis (PCA) is an unsupervised machine learning method that reduces high dimensional data without losing much information and also helps to identify hidden trends in the data. This technique has been applied in different astronomical problems such as spectral classification of stars (Storrie-Lombardi et al. \citeyear{Storrie-Lombardi1994}; Singh et al. \citeyear{Singh1998}), galaxies (Galaz $\&$ de Lapparent \citeyear{Galaz1998}), quasars (Yip et al. \citeyear{Yip2004}), analysis of galaxy velocity curves (Kalinova et al. \citeyear{Kalinova2017}), characterization of Type Ia Supernova light curves (He et al.   \citeyear{He2018}), etc. As par as the variable stars are concerned, Hendry et al. (\citeyear{Hendry1999}) found that Fourier decomposition encoded via PCA provides an efficient method to reconstruct noisy and sparse Cepheid light curves. Tanvir, Ferguson $\&$ Shanks (\citeyear{Tanvir1999}) applied PCA to statistically characterize V and I band of Hubble space telescope (HST) data for observed Cepheids in M96. In their paper, Kanbur et al. (\citeyear{Kanbur2002}) showed that PCA is more efficient than Fourier decomposition method at bringing out changes in light curve shape as a function of period. They also suggested that PCA will be more capable for searching resonances in observed Cepheid light as well as velocity curves. Leonard et al. (\citeyear{Leonard2003}) successfully applied the PCA approach to estimate mean magnitudes for HST observed Cepheids in NGC 1637. To analyse the structures of the RRab light curves using PCA, Kanbur $\&$ Mariani (\citeyear{Kanbur2004}) found that the correlation between principal components was significantly smaller than the correlation between Fourier amplitudes, leading to a more accurate estimate of absolute magnitudes. Tanvir et al. (\citeyear{Tanvir2005}) applied the PCA approach to confirm that Cepheids light curve shapes are different for Milky Way and the Large and Small Magellanic Clouds at the same period range. However, in the studies like Leonard et al. (\citeyear{Leonard2003}), Kanbur $\&$ Mariani (\citeyear{Kanbur2004}), Tanvir et al. (\citeyear{Tanvir2005}), the Fourier coefficients had been used as input for PCA analysis rather than the light curves themselves. In their paper, Deb $\&$ Singh (\citeyear{Deb2009}) used the original light curve data as input to the PCA since there is no advantage in the use of PCA where the input data be the Fourier coefficients. Meanwhile, they checked the relative performance of PCA as compared to the Fourier method for finding resonances in Cepheids and classification of different types of variable stars and concluded that the PCA technique could be applied to a larger and more diverse databases because of its efficiency and accuracy in a hierarchical classification. Recently, the PCA method has been applied in many studies including variable stars especially Cepheid variables at optical and near-infrared wavelengths (Bhardwaj et al. \citeyear{Bhardwaj2017}) and RR Lyrae light curves (Hajdu et al. \citeyear{Hajdu2018}).\\
Independent component analysis (ICA) is also an unsupervised machine learning algorithm similar to PCA, but the main focus is on the independence of components whereas PCA focuses on maximizing the variance of the data points. Recently, ICA method had been applied in many studies in modern astronomy such as source separation (Pires et al. \citeyear{Pires2006}; Sheldon $\&$ Richards \citeyear{Sheldon2018}), unsupervised classification (Das et al. \citeyear{Das2015}; Chattopadhyay et al.
\citeyear{Chattopadhyay2019}), dimension reduction (Sarro et al. \citeyear{Sarro2018}), etc. In their paper, Deb $\&$ Singh (\citeyear{Deb2009}) demonstrated the ability of PCA over the Fourier decomposition method in the case of classifying stars into different variability classes and this happened may be due to the fact that the principal components (PCs) are uncorrelated. However, in the case of ICA, the components are independent. In this sense, ICA might have some advantage over PCA, that is the motivation behind this paper.\\
In this paper, we collect huge amount of available light curve data sets of variable stars (Cepheids, RR Lyrae stars, eclipsing binaries, Mira variables)  mainly from OGLE-IV photometric databases for Large Magellanic Cloud (LMC), Small Magellanic Cloud (SMC) and  Milky Way (MW) galaxies and use PCA and ICA directly to the data and compare the relative performances and also compare with Fourier method used in the above studies.\\
The main features of this paper are following:
\begin{itemize}
   \item Structural analysis for Cepheids (FU, FO) classes of variables separately for LMC, SMC and MW using PCA and ICA both and comparison of the results with Fourier method mentioned in the literature.
    \item Classification of huge amount of data sets through PCA and ICA individually and to compare the results.
    \item Application of K-means clustering analysis on independent components (ICs) to plot the period-luminosity diagram as well as the colour-magnitude diagram (CMD).
\end{itemize}
The work has been arranged in the following manner. We present the data sets in Section \ref{sec:data}. The methods have been described in Section \ref{sec:STATISTICAL ANALYSES}. Results and discussions of structural analysis including classification and use of K-means Clustering have been narrated in Section \ref{sec: Results and Discussions}. Finally, we conclude the study in Section \ref{sec: Conclusions}.\\

\section{DATA SET} \label{sec:data}

The ultimate purpose of the Optical Gravitational Lensing Experiment (OGLE) is to search for dark matter using microlensing phenomena, introduced by Paczy{\'n}ski (\citeyear{Paczynski1986}). However, a wealth of information about variable stars has been collected for the Milky Way as well as nearby irregular galaxies, namely Large Magellanic Cloud (LMC) and Small Magellanic Cloud (SMC) as a by-product of this project.
The error in the photometric measurements of a variety of variable stars has been less ($\sim$ 0.006$-$0.2 mag), making the OGLE database a rich resource for analysing the properties of these stars. In this paper, we use an enormous amount of light curve data of RR Lyrae (RRab and RRc), Cepheids (fundamental mode (FU), first overtone (FO), Type$-$II), eclipsing binaries (EB) and long-period variable star like Mira variables separately for LMC, SMC and Milky Way galaxies, collected from the publicly available online database of OGLE\footnote{\url{https://ogle.astrouw.edu.pl/}} project.

\begin{table*}
\centering
\caption{Description of data sets collected from OGLE surveys}\label{tab:1}
\begin{tabular}{p{1.2cm}|p{1.2cm}|p{2.5cm}|p{0.5cm}|p{1.2cm}|p{2.5cm}|p{0.5cm}|p{1.2cm}|p{2.6cm}|p{0.5cm}}
 \hline \hline
 Data & \multicolumn{3}{|c|}{LMC}  & \multicolumn{3}{|c|}{SMC}  & \multicolumn{3}{|c|}{Milky Way}  \\
 \hline
 I$-$band & stars  \quad \quad selected & References & Set & Stars \quad \quad selected & References & Set & Stars \quad \quad selected & References  & Set\\
  \hline \hline
  
  RRab & 28,858 & Soszy{\'n}ski et al. (\citeyear{Soszynski2016}) & 1A & 5,193 &  Soszy{\'n}ski et al. (\citeyear{Soszynski2016}) & 2A & 19,568 & Soszy{\'n}ski et al. (\citeyear{Soszynski2011}) & 3A \\
  
   \cline{1-2} \cline{4-5} \cline{7-8} \cline{10-10}
   
  RRc & 9,903 & & 1B & 831 & & 2B & 7,231 & Soszy{\'n}ski et al. (\citeyear{Soszynski2019}) & 3B\\
  \hline
  
  FU & 2,428 & Soszy{\'n}ski et al. (\citeyear{Soszynski2015}) & 1C & 2,739 & Soszy{\'n}ski et al. (\citeyear{Soszynski2015}) & 2C & 1,184 & Soszy{\'n}ski et al. (\citeyear{Soszynski2020}) & 3C \\
  
 \cline{1-2} \cline{4-5} \cline{7-8} \cline{10-10}
 
  FO & 1,766 & & 1D & 1,783 & & 2D & 542 &  & 3D \\
 \hline
 
  Type$-$II & 286 & Soszy{\'n}ski et al. (\citeyear{Soszynski2018}) & 1E & 53 & Soszy{\'n}ski et al. (\citeyear{Soszynski2018}) & 2E & 1,374 & Soszy{\'n}ski et al. (\citeyear{Soszynski2020}) & 3E \\
  \hline
  
  EB & 26,121 & Graczyk et al. (\citeyear{Graczyk2011})  & 1F & 7,938 & Pawlak et al. (\citeyear{Pawlak2016})  & 2F & 17,590 & Soszy{\'n}ski et al. (\citeyear{Soszynski405})  & 3F \\
  \hline
  
  Mira & 1,663 & Soszy{\'n}ski et al. (\citeyear{Soszynski2009}) & 1G & 352 & Soszy{\'n}ski et al. (\citeyear{Soszynski217}) & 2G & 6,528 & Soszy{\'n}ski et al. (\citeyear{Soszynski2013}) & 3G \\
 \hline \hline
  
\end{tabular}
\end{table*}

For Milky Way, we collected both the galactic bulge and the galactic disk light curve data separately and add them to use in this analysis. The description of the data set including references is included under Table \ref{tab:1}.

\section{Methodology and  Statistical Analyses} \label{sec:STATISTICAL ANALYSES}

\subsection{Conversion of data sets into matrix form} \label{subsec:Conversion of datasets into Matrix form}

In this paper, we have used enormous amount of I-band light curve data collected from recent OGLE surveys for different variable stars like RR Lyrae, Cepheids, Mira, Eclipsing binary, etc. in LMC, SMC, and Milky Way galaxies to analyse the structural patterns as well as classification of these stars. Since the input to PCA or ICA has to be a matrix, we first convert each of the data sets into matrix form. Each of the light curve data file mainly consists of Heliocentric Julian Days as observational times, observed I-band magnitudes and uncertainty of magnitudes. The conversion procedure is structured below.

\begin{figure}
\includegraphics[width=9cm,height=7cm]{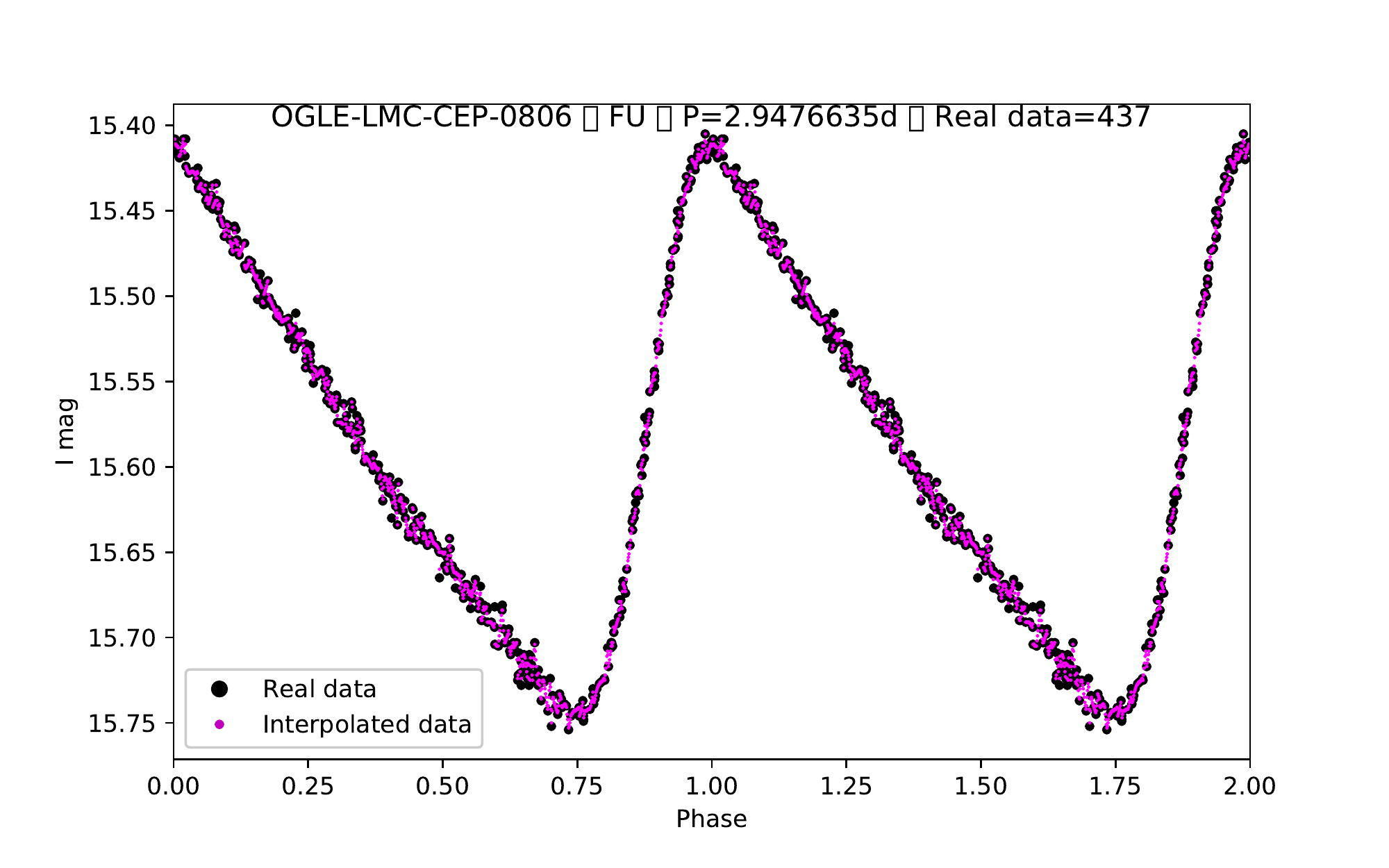}
\caption{Spline interpolation of LMC FU light curve to obtain 1000 I-band magnitudes.}\label{fig:1}
\end{figure}

\begin{itemize}
    \item Convert the observation time (t) to phase ($\phi$) with the help of equation (\ref{eqn:1}).
    \begin{equation}\label{eqn:1}
        \phi = \frac{(t-t^\prime)}{P} - int\bigg[\frac{(t-t^\prime)}{P}\bigg]
    \end{equation}
    \hfill (Ngeow et al.\citeyear{Ngeow2003}; Deb $\&$ Singh \citeyear{Deb2009})\\
    where $\phi$ ranges from 0 to 1, corresponds to one full cycle of pulsation and $int$ refers to the integer part of the content. P and $t^\prime$ are respectively the period (in days) and epoch of the maximum light, taken from literature of the respective databases referred in Table \ref{tab:1}.
    \item Interpolate magnitudes of each light curve using spline interpolation method between phase 0 to 1 with step length 0.001 to obtain 1000 magnitudes. Mathematically, linear splines are fitted at n equally spaced phase intervals to give tabulated function 
    \begin{equation}
        z_i = z(y_i), \quad i=1,2,3,...,n_j
    \end{equation}
    with j being the jth light curve. Therefore, for i=1,2,3,...,${n_j}$ we get $({n_j}-1)$ linear functions of the form
    \begin{equation}
        f_i(y) = p_iz_i+q_iz_{i+1}, \quad  y \in [y_i,y_{i+1}]
    \end{equation}
    where $f_i(y_i)=z_i$ and $f_i(y_{i+1})=z_{i+1}$. 
    The constants $p_i$ and $q_i$ satisfies
    \begin{equation}
        p_i = \frac{y_{i+1}-y}{y_{i+1}-y_i} \quad and \quad q_i = 1-p_i = \frac{y-y_i}{y_{i+1}-y_i} ,
    \end{equation}
    i=1,2,3,...,$(n_j-1)$. \\
    We repeat the process with newly formed interpolating points and old ones together until obtaining desired numbers. Thus interpolation error decreases by following the above procedure (Cassisi et al. \citeyear{Cassisi2012}).\\
    Example of one of the interpolated light curves for variable stars has been demonstrated in Fig. \ref{fig:1}.

\item After interpolating all the light curves corresponding to a data set, we have normalized the magnitudes of each light curve between 0 and 1 using equation (\ref{eqn:5}).
\begin{equation}\label{eqn:5}
  m_{norm} = \frac{m-m_{min}}{m_{max}-m_{min}}
 \end{equation}

\item Finally, we organize the normalized magnitude data into the matrix form of order M$\times$1000, where M is the number of light curves for further analysis. We have used a high-level programming language in  Python for our analysis.
\end{itemize}
 
\subsection{Principal component analysis (PCA)} \label{subsec:Principal Component Analysis (PCA)}
In the present work, PCA has been used in two different ways. Firstly, to dig out the hidden structural changes from the variable star data and to classify between different variable stars with respect to their periods. In another case, PCA has been performed prior to ICA to obtain optimum number of ICs.\\
PCA is basically a way of visualizing hidden patterns of complex high dimensional (say k) data sets by transforming the original data sets orthogonally into new set of variables (principal components) of fewer dimensions (r $< <$ k) which basically retain most of the  variation in the original data. The method has been described below in short.\\
Let us consider a matrix $M_{k\times q}$ consisting of $l_{ij}$'s, be the q normalized magnitudes of k light curves. We perform standardization on the said matrix using the following equation 
\begin{equation}
    r_{ij} = \frac{l_{ij}-\overline l_j}{s_j\sqrt{k}}
\end{equation}
where $r_{ij}$'s are the elements of standardized matrix $T_{k\times q}$ and 
\begin{equation}
   \overline l_j = \frac{1}{k} \sum_{i=1}^{k} l_{ij}
\end{equation} 
\begin{equation}
    s_j^2 = \frac{1}{k} \sum_{i=1}^{k} (l_{ij}- \overline l_j)^2
\end{equation}
are the mean value and standard deviation of any element of the matrix $M_{k\times q}$. The corresponding Correlation matrix D is given by 
\begin{equation}
    D_{jh}= \sum_{i=1}^{k} r_{ij} r_{ih} 
          = \frac{1}{k} \sum_{i=1}^{k} (l_{ij}- \overline l_j)(l_{ih}- \overline l_h) / (s_j s_h)
\end{equation}
The principal components are the eigen vectors ($e_i$) corresponding to the eigenvalues ($\lambda_i$) of the equation
\begin{equation}
   D {e_i} = {\lambda_i} {e_i} 
\end{equation}
The eigen vector associated with the largest eigenvalue corresponds to the axis of maximum variance and is called the first principal component. Similarly, the second principal component is the eigenvector associated with the second largest eigenvalue and it corresponds to the axis orthogonal to first and so on. The variance-ratio explained by the $j^{th}$ eigenvalue is given by  $\lambda_j/S$, where S be the sum of eigenvalues (Singh et al. \citeyear{Singh1998}).  

\subsection{Independent component analysis (ICA)} \label{subsec:Independent Component Analysis (ICA)}
Independent component analysis (ICA) is also a technique which reduces the dimension of a higher dimensional matrix as similar to PCA but the primary difference is that the components are mutually independent in the former case.\\
Mathematically, let, $Y_1, Y_2,..., Y_k$ be k random vectors (here light curves) and q (here 1000) be the number of magnitudes of each $Y_i$, (i$=$1,2,3,...,k).\\
Let, Z $=$ BQ, where $Q = [Q_1,Q_2,...,Q_k]^\prime$ is a random vector consisting of hidden components $Q_i$, (i=1,2,3,...,k) such that $Q_i$'s are mutually independent and B is non singular matrix. Then,
\begin{equation} \label{eqn:11}
    Q = B^{-1}Z = WZ
\end{equation}
Where W is the unmixing matrix (Comon \citeyear{Comon1994}; Chattopadhyay et al. \citeyear{Chattopadhyay2013}, and references therein). The main objective of ICA is to find Q with the help of W such that $cov (g_1(Q_i),g_2(Q_j))$ $=$ 0 for i $\neq$ j, where $g_1(Q_i), g_2(Q_j)$ are any two non linear functions.\\
Choice of optimum number of ICs is difficult (Kairov et al. \citeyear{Kairov2017}). In this work, we use PCA to choose optimum number of ICs. The optimum number of PCs are choosen in a way that the total variance obtained by PCs are above 85$\%$ and the choice of optimum number of ICs are based on choice of optimum number of PCs (Albazzaz and Wang \citeyear{Albazzaz2004}; Chattopadhyay et al. \citeyear{Chattopadhyay2013}; Chattopadhyay et al. \citeyear{Chattopadhyay2019}). Other choices of optimum number of ICs leads to unsatisfactory results as being found by multiple run of the algorithm. For PCA, PCs contained variance of the data set in a decreasing order i.e. PC1 contained maximum variance, next PC2 contained less variance than PC1 and so on, however in the case of ICA, there is no definite pattern for ICs for which the variance in the data set is maximum. As the variation among ICs has no order like PC components so by visualizing the plots for individual ICs corresponding to each data set, we have selected that component of ICA which has the maximum variation for comparison with PC1, which has maximum variation. So the independent component with maximum variation may be different for different data sets. For example, in Fig. \ref{fig:5}, we plot PC1 against log P (in days) and IC3 against log P (in days) as PC1, IC3 both have maximum variance in their respective analysis. Here we have used scikit-learn machine learning library for Python programming for both PCA and ICA.

\subsection{K-means cluster analysis} \label{K-means cluster analysis}
K-means cluster analysis (CA) is a  multivariate unsupervised machine learning technique, basically applied for finding coherent groups in a data set (Chattopadhyay et al. \citeyear{Chattopadhyay2007}, \citeyear{Chattopadhyay2010}, \citeyear{Chattopadhyay2012}, \citeyear{Chattopadhyay2019}; Das et al. \citeyear{Das2015}). The main objective of this method is to find K clusters such that all the objects are distributed among those K clusters with the following two properties :\\
\textbullet\ Each cluster must have at least one object.\\
\textbullet\ Each object must be contained in exactly one cluster.\\

\begin{figure}
\includegraphics[width=8cm,height=6cm]{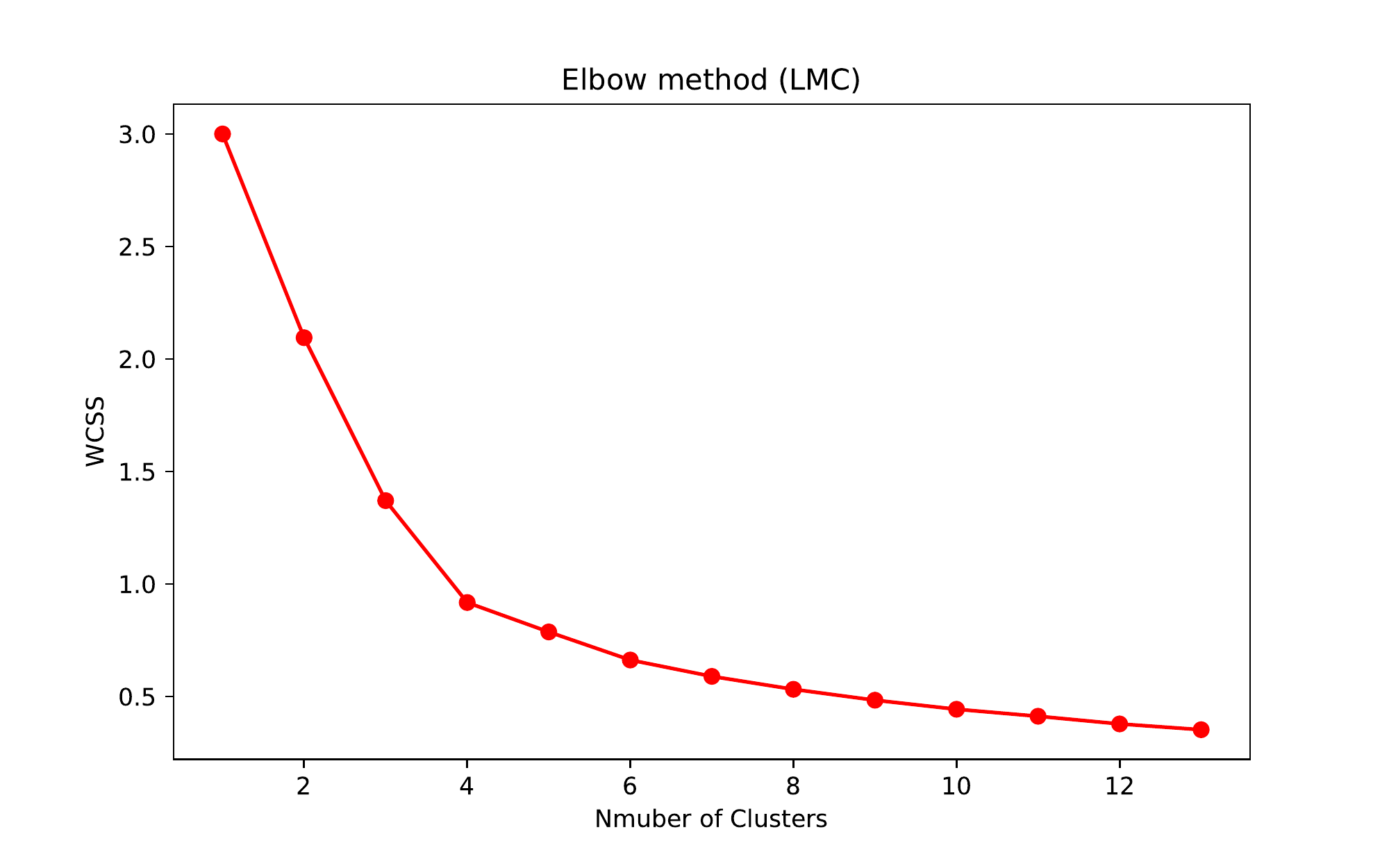}
\caption{Representation of within cluster sum of squares (WCSS) against number of clusters to determine optimum value of K. For LMC, the optimum value of clusters (K) is 4.}\label{fig:2}
\end{figure}

The algorithm of this method is described as below :\\

\noindent Step 1: All objects are randomly distributed among K (given) groups.\\

\noindent Step 2:  A particular object is selected randomly from any of those groups at first. Then parametric distance (parameters are the number of ICs choosen for the analysis) is calculated between the selected object and the remaining objects. If the distance measured between objects in the group is greater than that for objects in other groups, they are alternated.\\

\noindent Step 3: The previously described procedure is applied for all objects and for all groups.\\

\noindent Step 4: Repeat steps 1 and 3 until no further change of objects between groups.

Since the number of clusters (K) is pivotal input to the above algorithm, it is crucial to find K accurately. Thus we opted for Elbow method described below.

\subsubsection{Determination of optimum value of K}

The optimum number of clusters is selected by K-means clustering (MacQueen \citeyear{MacQueen1967}) in association with the method proposed by Sugar $\&$ James (\citeyear{Sugar2003}). In this method, distortion measure ($d_k$) is calculated for K $=$ 1,2,3... etc with the help of equation (\ref{eqn:12}),
\begin{equation} \label{eqn:12}
   d_k = (1/P) min_x E[(x_k - c_k)^\prime (x_k - c_k)],
\end{equation}
where P is the total number of variables (here the number of  ICs). In other words, it is the distance of $x_k$ vector (values of parameters) from centriod $c_k$ of the corresponding cluster. The optimum value of K is selected for which $d_k$ (within cluster sum of squares, WCSS) first starts to diminish at the `elbow' of the pattern. For SMC, LMC (see Fig. \ref{fig:2}) and MW, we found K $=$ 4.

\section{Results and Discussions} \label{sec: Results and Discussions}

\subsection{Identifying resonances for Cepheid light curves: a comparison between PCA and ICA }

The usage of Cepheid variables are profound in various astrophysical investigations such as determination of Hubble's constant to a certain accuracy (Freedman $\&$ Madore \citeyear{Freedman2010}; Riess et al. \citeyear{Riess2018}) as well as extragalactic distance scaling through period-luminosity (P-L) relation (Leavitt and Pickering \citeyear{Leavitt1912}). Also, more specifically classical Cepheids are being utilized as the convenient touchstone for verification of stellar pulsation theories (Antonello \citeyear{Antonello1994}; Bono et al. \citeyear{Bono1999}). Radial modes of pulsation i.e. periodic expansion and contraction in outer layers of Cepheids lead to the variations of stars luminosity. These inward-outward movements might be due to opacity driven mechanism. In the dim phase, due to a sufficiently hot environment helium become fully ionized which causes an increase in opacity of the entire layer and absorbs radiation outpouring from the core of the star, leading to the expansion of the star's radius. This expansion triggers cooling of the outer layer which helps to recombine the electrons of helium to produce single ionized helium, which reduces the opacity as well as the ability of absorption, causing a brighter appearance of the star as well as contraction of the layers and this contraction provokes the next expansion. Fundamental mode is the simplest radial mode in which the expansion/contraction happens simultaneously to the entire environment of the star whereas in the case of the first overtone, while the outer zone of the star is expanding, the inner zone shrinks and vice versa.\\
Systematic changes in the shape of classical Cepheid light curves arranged according to increasing pulsation period is known as Hertzsprung progression (HP) (Hertzsprung \citeyear{Hertzsprung1926}) .There is a secondary bump appears in the descending branch for light curves of short period (P $<$ 10 days) Cepheids and as the period increases the bump moves backward in phase and appears at the maximum light, called the centre for HP  and finally the bump moves to the rising branch of long period (P $>$ 10 days) Cepheids. This phenomenon is associated with the sharp breaks found around 10 days in the progression of Fourier parameters against period  (Simon $\&$ Lee \citeyear{Simon1981}) and these could be the consequence of a resonance between the fundamental mode and second overtone Cepheids when the period ratio between the modes is $\sim$ 0.5 (Simon $\&$ Schmidt \citeyear{Simon1976}; Simon \citeyear{Simon1977}; Takeuti $\&$ Aikawa \citeyear{Takeuti1981}; Aikawa \citeyear{Aikawa1984}; Klapp et al. \citeyear{Klapp1985}).
Since then the well known FD method had been used by many authors (Simon $\&$ Lee \citeyear{Simon1981};  Moskalik et al. \citeyear{Moskalik1992}; Welch et al. \citeyear{Welch1997}; Beaulieu \citeyear{Beaulieu1998}) to identify the possible resonances in FO and FU Cepheids as well as  determination of centre of HP by the means of Fourier parameters for different galaxies. By finding out the center of HP in LMC (Z = 0.008), SMC (Z = 0.004) and MW (Z = 0.02), Beaulieu (\citeyear{Beaulieu1998}) concluded that a decrease in metallicities shifts the HP  center along longer periods. Bono et al. (\citeyear{Bono2000}) constructed nonlinear convective models by adopting a fixed chemical composition (Y = 0.25, Z = 0.008) to investigate bump Cepheids theoretically and showed that the periods of the HP center are in good agreement with observational values. They also suggested that the dependence of the HP behaviour on chemical compositions should be investigated with the help of detailed comparison between  theoritical nonlinear Cepheid models and observational evidences. Thus reliable detection of resonances for  Cepheids through various modern tools is crucial as they can be used to place constraints on stellar parameters such as luminosity, mass, metallicity etc. (Simon $\&$ Kanbur \citeyear{Simon1995}). However, FD method is time consuming in case of large databases as each light curve has to be analysed separately. In recent years, PCA has been widely used as an alternative to FD method in identifying resonances along the light curve shape with the help of vastly available observed light curve data (Kanbur et al. \citeyear{Kanbur2002};  Kanbur $\&$ Mariani \citeyear{Kanbur2004}; Tanvir et al. \citeyear{Tanvir2005}; Deb $\&$ Singh \citeyear{Deb2009}). ICA is also a dimension reduction technique similar as PCA and hopefully it can also contribute to identify structural changes around FU and FO Cepheids. This is the motivation behind this part.
In this section, the capability of both PCA and ICA have been compared in the case of extracting resonances of FU and FO Cepheids individually for LMC, SMC and MW along with FD method mentioned in the paper Deb $\&$ Singh (\citeyear{Deb2009}). 

\begin{figure*}  
\includegraphics[width=18cm,height=5cm]{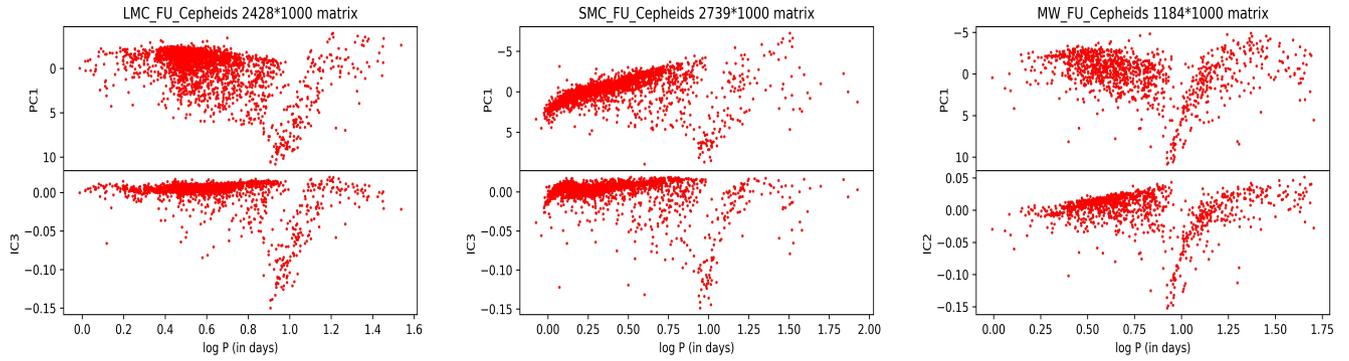}
\caption{Components of PCA and ICA have been plotted as a function of log P (in days) separately for the 2428 I-band FU Cepheids in LMC (left column), 2739 I-band FU Cepheids in SMC (middle column) and 1184 I-band FU Cepheids in MW (right column). Upper panel is for PCA component and lower panel is for ICA component in each of the columns}\label{fig:3}
\end{figure*}

\begin{figure*}
\includegraphics[width=18cm,height=5cm]{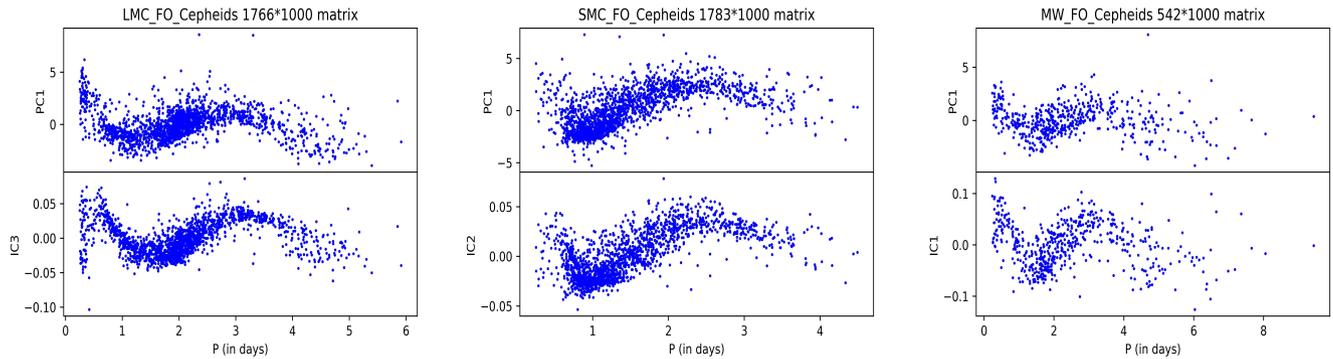}
\caption{Components of PCA and ICA have been plotted as a function of P (in days) separately for the 1766 I-band FO Cepheids in LMC (left column), 1783 I-band FO Cepheids in SMC (middle column) and 542 I-band FO Cepheids in MW (right column). Upper panel is for PCA component and lower panel is for ICA component in each of the columns.}\label{fig:4}
\end{figure*}

\subsubsection{Fundamental mode (FU) Cepheids}

All the light curve data have been analysed for FU Cepheids are from latest OGLE-IV database. Deb $\&$ Singh (\citeyear{Deb2009}) used data from various sources (mostly from OGLE) and add LMC and SMC (I-band and V-band) data altogether for analysis, but we have used only I-band data separately for LMC, SMC and MW (data sets 1C, 2C, 3C respectively from Table \ref{tab:1}). For LMC, the input matrix consisting of 2428 $\times$ 1000 array corresponding to 2428 light curves for FU Cepheids and each consists of 1000 data points (here magnitudes) between phase 0 to 1 with step length 0.001. Table \ref{tab:2} explained about output after performing PCA on 2428 $\times$ 1000 data matrix.

\begin{table}
\centering
\renewcommand{\arraystretch}{1.3}
\caption{First 3 eigenvalues, their explained variance and cumulative explained variance for 2428 FU Cepheids}\label{tab:2}
\begin{tabular}{@{} p{0.4 cm} p{1.9 cm} p{1.9 cm} p{2.7 cm}}
\hline 
 PC & Eigenvalue ($\lambda$) & Percentage & Cum. Percentage  \\
\hline
1 & 5.615142 &  55.9325 &  \quad  55.9325 \\
2 & 2.348315 &  23.3916 &  \quad  79.3241 \\
3 & 0.944975 &   9.4129 &  \quad  88.7370 \\
\hline
\end{tabular}
\end{table}

The first 3 PCs (i.e. $ \lambda \geq 1$) contain more than 85$\%$ of the  variance in the data. Thus we consider the optimum number of ICs as 3 for LMC. Similarly by performing PCA on a data matrix of order 2739 $\times$ 1000 corresponding to 2739 light curves for the  SMC FU Cepheids, we found that first 4 PCs (i.e. $\lambda \geq 1$) explained above 85$\%$  of the variance in the data. Thus for SMC, optimum number of ICs is equal to 4. Finally, for MW FU Cepheids (1184 $\times$ 1000), almost 85$\%$ of the variance explained by 4 PCs (i.e. $ \lambda \geq 1$). Therefore in this case the optimum number of ICs is equal to 4. Components correspond to PCA and ICA are plotted against log P (in days) in Fig. \ref{fig:3}. Deb $\&$ Singh (\citeyear{Deb2009}), in their paper found a discontinuity in the shape around log P $=$ 1 by applying FD and PCA techniques separately for LMC FU Cepheids however the discontinuity is more pronounced for the former one. It is evident from Fig. \ref{fig:3} that the discontinuity around log P $=$ 1 is more prominent for ICA component compared to PC1  although we have used more light curve data for LMC as well as SMC and  MW FU Cepheids in our analysis.
Due to lack of data points, Kanbur et al. (\citeyear{Kanbur2002}) could not find any resonance suggested by Antonello $\&$ Morelli (\citeyear{Antonello1996}) in the period range 1.38 $<$ log P $<$ 1.43 using PCA on Fourier coefficients. However, structural changes around log P $\sim$ 1.5 are visible both for PCA and FD analysed by Deb $\&$ Singh (\citeyear{Deb2009}). Using PCA on a larger data set of LMC classical Cepheids, structural change around log P $\sim$ 1.35 is visible (left column in Fig. \ref{fig:3}) but not so clear because of scattering of data points in the period range 1.2 $<$ log P $<$ 1.6. On the contrary in the case of ICA, the structural pattern is much more precise and  the location of change is  relatively clear which is log P $\sim$ 1.3, close to the range suggested by Antonello $\&$ Morelli (\citeyear{Antonello1996}).
 For SMC FU Cepheids, due to scattering of data points, the structural pattern beyond log P $=$ 1 is not clear for PC1 but the structural changes  around the period range 1.3 $<$ log P $<$ 1.4 is prominent in the case of IC3 which is almost similar to the range suggested by Antonello $\&$ Morelli (\citeyear{Antonello1996}). Also, there is slight change in pattern somewhere around log P $\sim$ 1.8 is visible for both the cases in SMC but confirmation of the existence of such resonance is difficult due to lack of data points in these period range. In such scenarios, radiative hydrodynamical models are useful (Kienzle et al. \citeyear{Kienzle1999}). In the case of MW Cepheids (right column in Fig. \ref{fig:3}), performance of ICA and PCA are similar.

\begin{figure*}
\includegraphics[width=18cm,height=8cm]{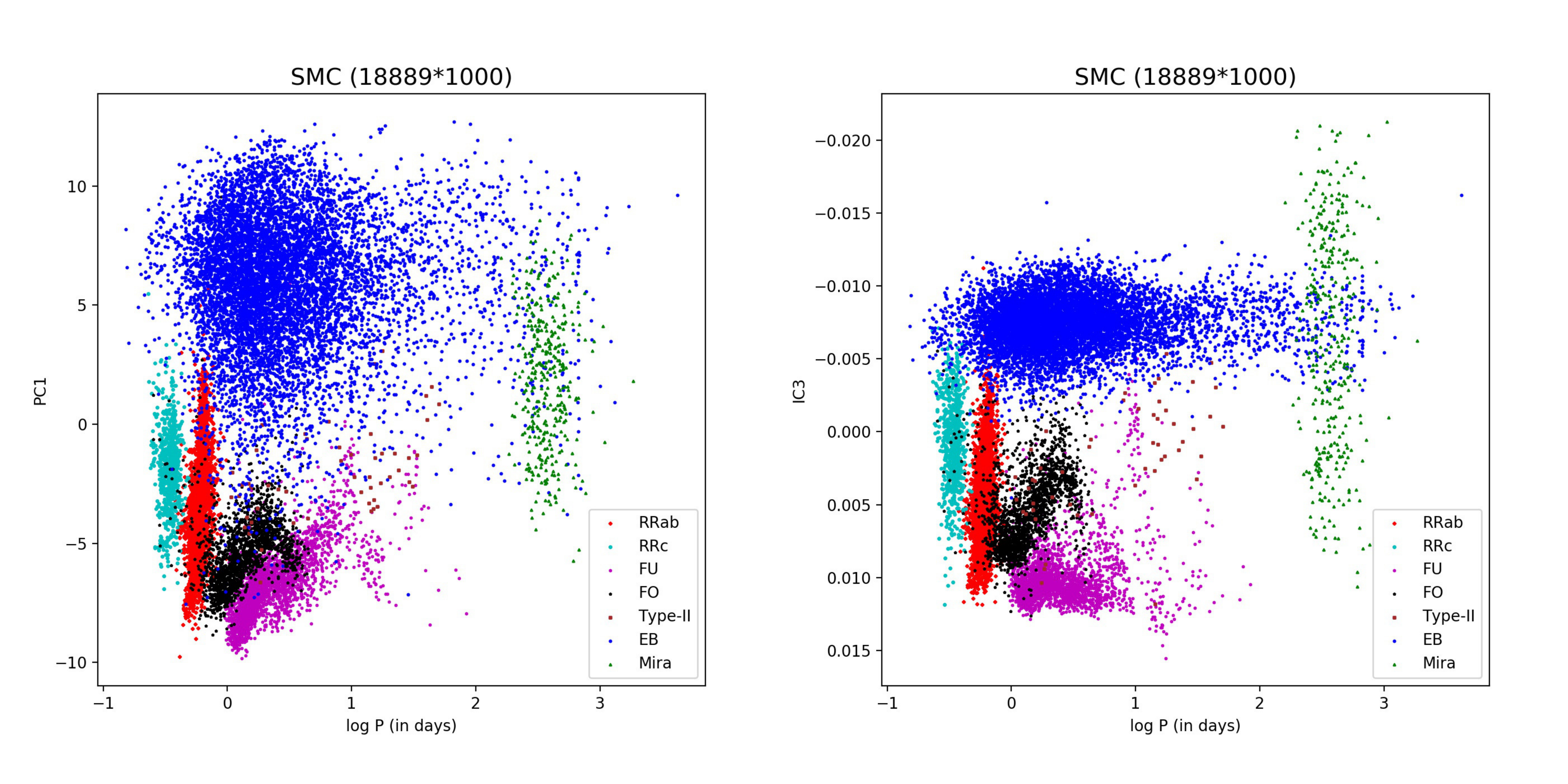}
\caption {Classification of different types of variable light curves based on PCA and ICA method separately. The input matrix is of order 18889$\times$1000 for SMC. PC1 plotted as a function of log P (in days) (left) and IC3 plotted against log P (in days) (right). Types of variable stars are outlined in plot legend.}\label{fig:5}
\end{figure*}

\subsubsection{First overtone (FO) Cepheids}
We have collected 1766 (set 1D in Table \ref{tab:1}), 1783 (set 2D in Table \ref{tab:1}) and 542 (set 3D in Table \ref{tab:1}) I-band light curve data of FO Cepheids towards LMC, SMC and MW respectively from OGLE-IV database. For LMC, first 3 PCs (i.e. $\lambda \geq 1$) corresponds to 85$\%$ of the variance in the data. So, we set optimum number of ICs as 3 for our analysis. Performing PCA on a 1783 $\times$ 1000 matrix for SMC, first 4 PCs (i.e. $ \lambda \geq 1$) contain approximately 85$\%$ variance. So, for SMC optimum number of ICs is taken as 4. However for analysis on MW FO Cepheids, it takes 5 PCs (i.e. $\lambda \geq 1$) to counter nearly 85$\%$ variance in the data. Thus we have taken 5 as the optimum number of components for ICA. Fig. \ref{fig:4} carries outputs for PCA and ICA respectively for LMC (left column), SMC (middle column) and MW (right column) data sets. A short period discontinuity emerges around P $\sim$ 0.4 days (similar result found by  Soszy{\'n}ski et al. (\citeyear{Soszynski2008}), Deb $\&$ Singh (\citeyear{Deb2009}) for LMC) for all the sub figures of Fig. \ref{fig:4}. For LMC FO Cepheids there is a change in shape around P $\sim$ 1.5 days, evident for IC3 but difficult to visualize in the case of PC1. Also, we found a definite pattern change around a period of $\sim$ 3.2 days for both PCA and ICA analyses could be explained by the occurrence of resonance between first and fourth overtone (Antonello $\&$ Poretti \citeyear{Antonello1986}; Peterson \citeyear{Peterson1989}). However, authors like Kienzle et al. (\citeyear{Kienzle1999}) and Feuchtinger et al. (\citeyear{Feuchtinger2000}) using hydrodynamical models of FO Cepheids have shown that 3.2 days is not a resonance but Deb $\&$ Singh (\citeyear{Deb2009}) found this as resonance using FD and PCA both. There is also some change visible around $\sim$ 4.5 days only in the case of PC1 suggested by both Buchler et al. (\citeyear{Buchler1996}) and  Kienzle et al. (\citeyear{Kienzle1999}). However ICA fails to detect such resonance similar as for FD analysed by Deb $\&$ Singh (\citeyear{Deb2009}). For SMC FO Cepheids, there is clearly a change in shape in period-luminosity relation around $\sim$ 2.5 days (Kanbur et al. \citeyear{Kanbur2002}; Bharadwaj et al. \citeyear{Bhardwaj2016}) visible for both PC1 and IC2. For MW FO Cepheids, the structural changes are visible around 1.5 days,  3.2 days. However IC1 does a better job by bringing out the changes than PC1. ICA performs better in bringing out the resonance around $\sim$ 4.5 days than PCA.\\
Thus in general, the performance of ICA technique is better than both PCA and FD methods to extract the structural patterns of FU and FO Cepheids for LMC, SMC and MW  respectively.

\begin{figure*} 
\includegraphics[width=18cm,height=8cm]{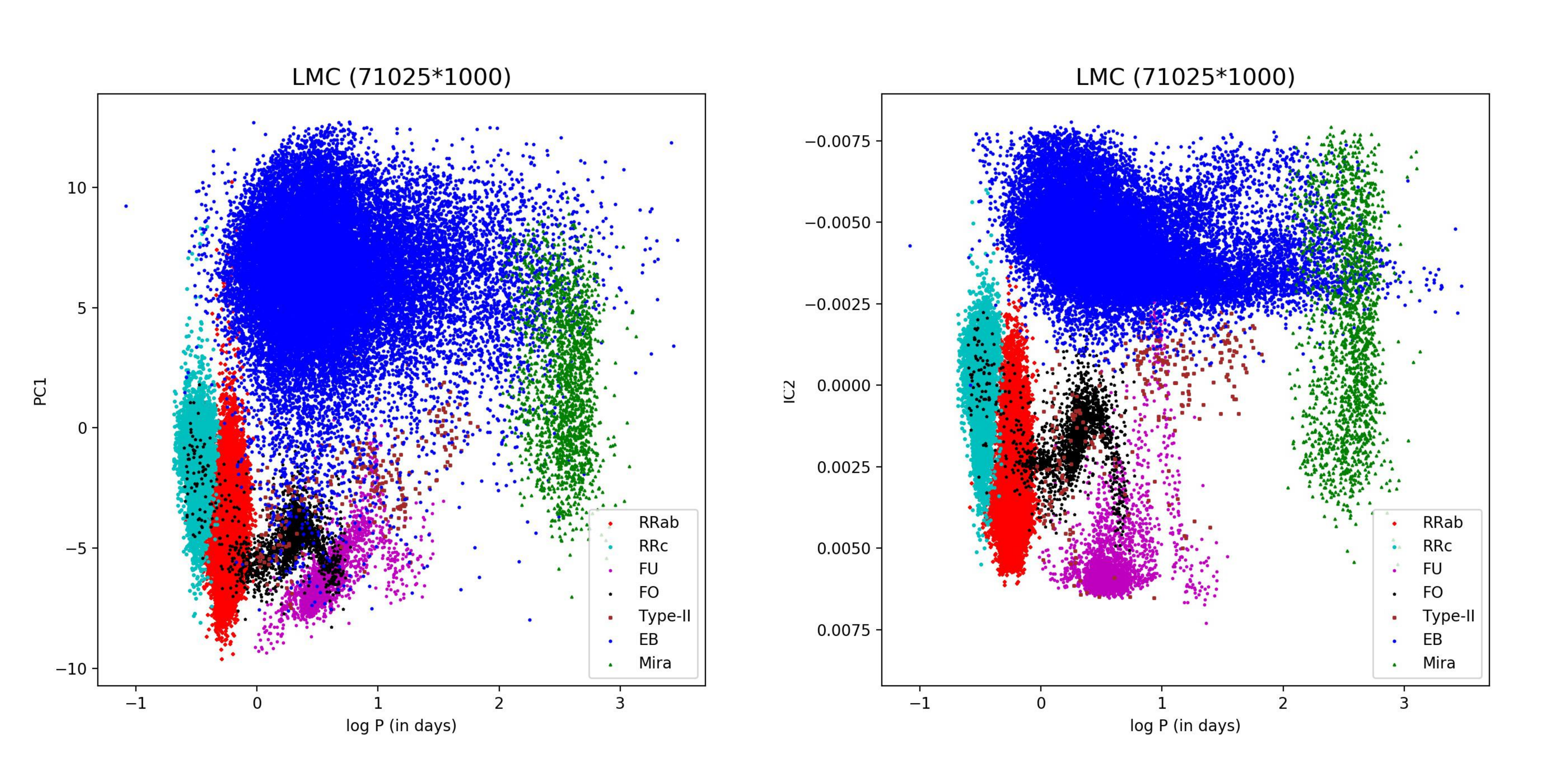}
\caption{Classification of different types of variable light curves  based on PCA and ICA method separately. The input matrix is of order 71025$\times$1000 For LMC. PC1 plotted as a function of log P (in days) (left) and IC2 plotted against log P (in days) (right). Types of variable stars are outlined in plot legend.}\label{fig:6}
\end{figure*}

\begin{figure*} 
\includegraphics[width=18cm,height=7.5cm]{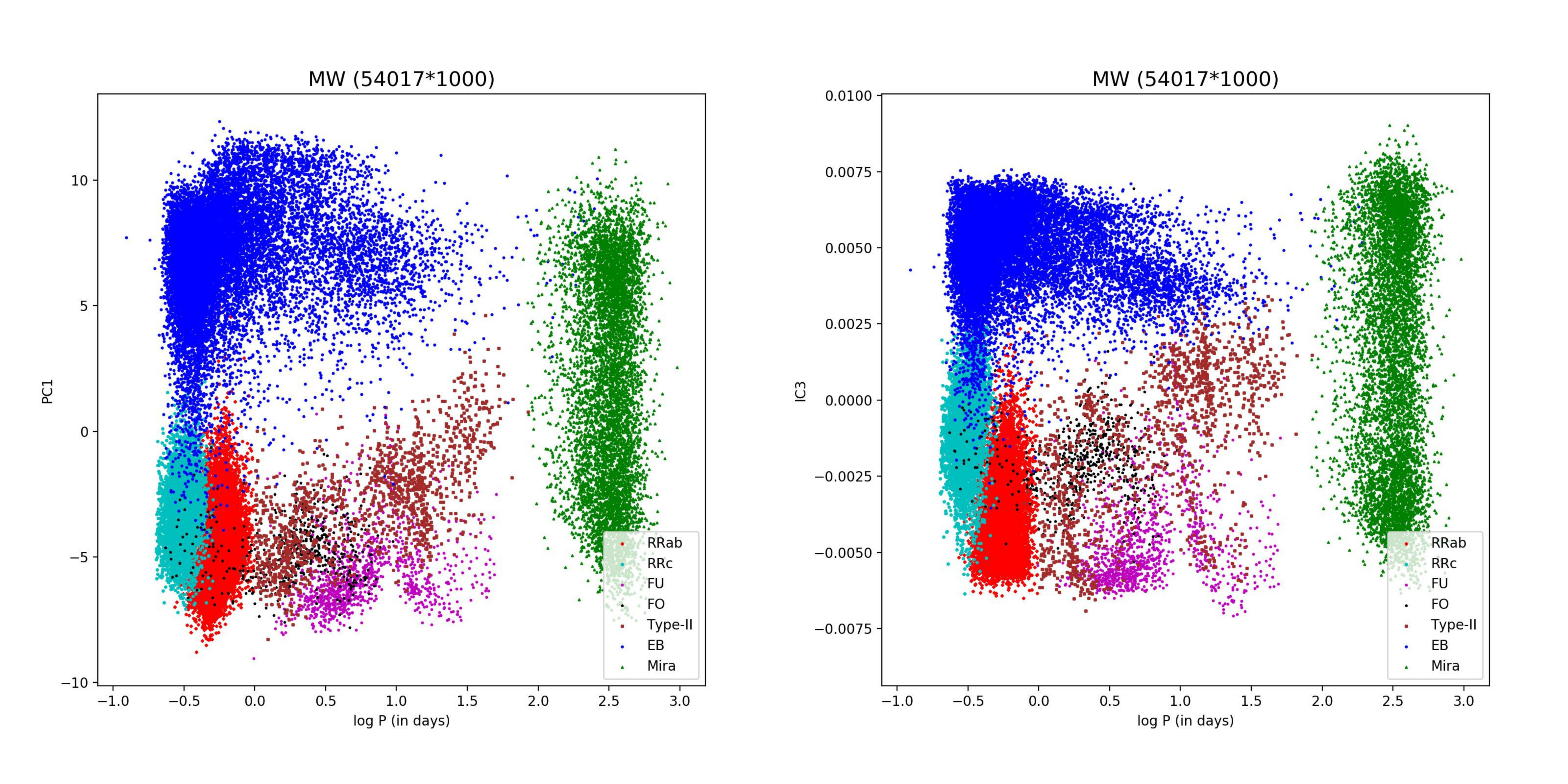}
\caption{Classification of different types of variable light curves based on PCA and ICA method separately. The input matrix is of order 54017$\times$1000 For SMC. PC1 plotted as a function of log P (in days) (left) and IC3 plotted against log P (in days) (right). Types of variable stars are outlined in plot legend.}\label{fig:7}
\end{figure*}

\subsection{Identification of PCA and ICA components : a comparative study}

In this section, we demonstrate the capabilities of unsupervised machine learning techniques like PCA and ICA in classifying different kinds of variable stars light curve data sets collected from OGLE database for LMC, SMC as well as MW, described in Table \ref{tab:1}. For SMC, PCA has been performed on a total of 18889 (set $2A+2B+2C+2D+2E+2F+2G$ in Table \ref{tab:1}) light curves containing 1000 magnitudes each. We found that the first 6 PCs (i.e. $\lambda \geq 1$) explain more than 85$\%$ of variance in the data set whereas PC1 contains 63.86$\%$. Thus the optimum number of ICs to be chosen as 6. PC1 and IC3 are plotted as a function of log P (in days) separately in Fig. \ref{fig:5} for all types of variable stars. There is some overlapping between eclipsing binaries (7938 in total) and other variable classes through out the period range for PC1 but the separation is much more efficient in the case of IC3, however some overlapping between eclipsing binaries and Mira variables are visible in the period range 2.5 $<$ log P $<$ 3, that might be due to some outliers  present in the data set of eclipsing binaries. Slight overlapping between RRab and FO Cepheids are seen in both cases (PCs and ICs) over a small period range. IC3 not only is able to separate FU and FO Cepheids to a large extend and  perform better than PC1 but also prominently figure out the discontinuity of FU variable around log P $=$ 1. PC1 and IC3 both fail to separate FO and Type-II Cepheids in 0 $<$ log P $<$ 1. However no overlapping is visible between FU and Type-II Cepheids. \\
For LMC the input matrix is 71025$\times$1000 which is basically a collection of 1000 interpolated magnitude data points having phase 0 to 1 with step length 0.001 each for 71,025 light curves (set $1A+1B+1C+1D+1E+1F+1G$ in Table \ref{tab:1}). We found out that the first 5 PCs (i.e. $\lambda \geq 1$) explain above 85$\%$ of variance in the data set whereas PC1 contains 62.36$\%$. Thus the optimum number of ICs to be taken is 5. Deb $\&$ Singh (\citeyear{Deb2009}) used Fourier decomposition parameter $R_{21}$ and PC1 to classify nearly about 17,000 stars and found that the sample of RR Lyraes (RRab and RRc) and Cepheids (FU and FO) could not separate well in the whole data set. From Fig. \ref{fig:6}, it can be seen that the amount of overlap between FO and RR Lyrae (RRab and RRc) samples is trivial in terms of sample data used for our analysis, however PC1 is not able to separate FU and FO Cepheids well enough which is not the case for IC2. Also, the discontinuity of FU Cepheids around log P = 1 is clearly visible for IC2. There is certain mix up between eclipsing binaries with other variable stars for PCA analysis whereas in the period range of 1 to 100 days, eclipsing binaries have distinct IC2 values from other variable stars. Considerable amount of overlapping between eclipsing binaries and Mira variables have been seen in both cases in the range of 2 $<$ log P $<$ 3, that might be due to presence of some outliers in the data set. Type-II variables are differentiated well for IC2 in the period range 1 $<$ log P $<$ 2 and 0$<$ log P $<$ 0.2. There is some overlapping between Type-II and FO for both the cases but in case of IC2 it is much reduced.\\
For MW galaxy, the input matrix is of order 54017 $\times$ 1000 (set $3A+3B+3C+3D+3E+3F+3G$ in Table \ref{tab:1}). Applying PCA, we figured out that the first 6 PCs (i.e. $\lambda \geq 1$) explain above 85$\%$ of variance in the data whereas PC1 contains 51.57$\%$. Thus for MW, the optimum number of ICs to be taken as 6. From Fig. \ref{fig:7}, it may be noted that samples of EB, RRab, RRc and Mira variables are well separated more for IC3 than PC1 though there are some overlapping among Cepheids FU, FO and Type-II. Overall it  is clear that the performance of ICA technique is much better for classification of big data than PCA and FD methods.

\begin{figure*} 
\includegraphics[width=18cm,height=10cm]{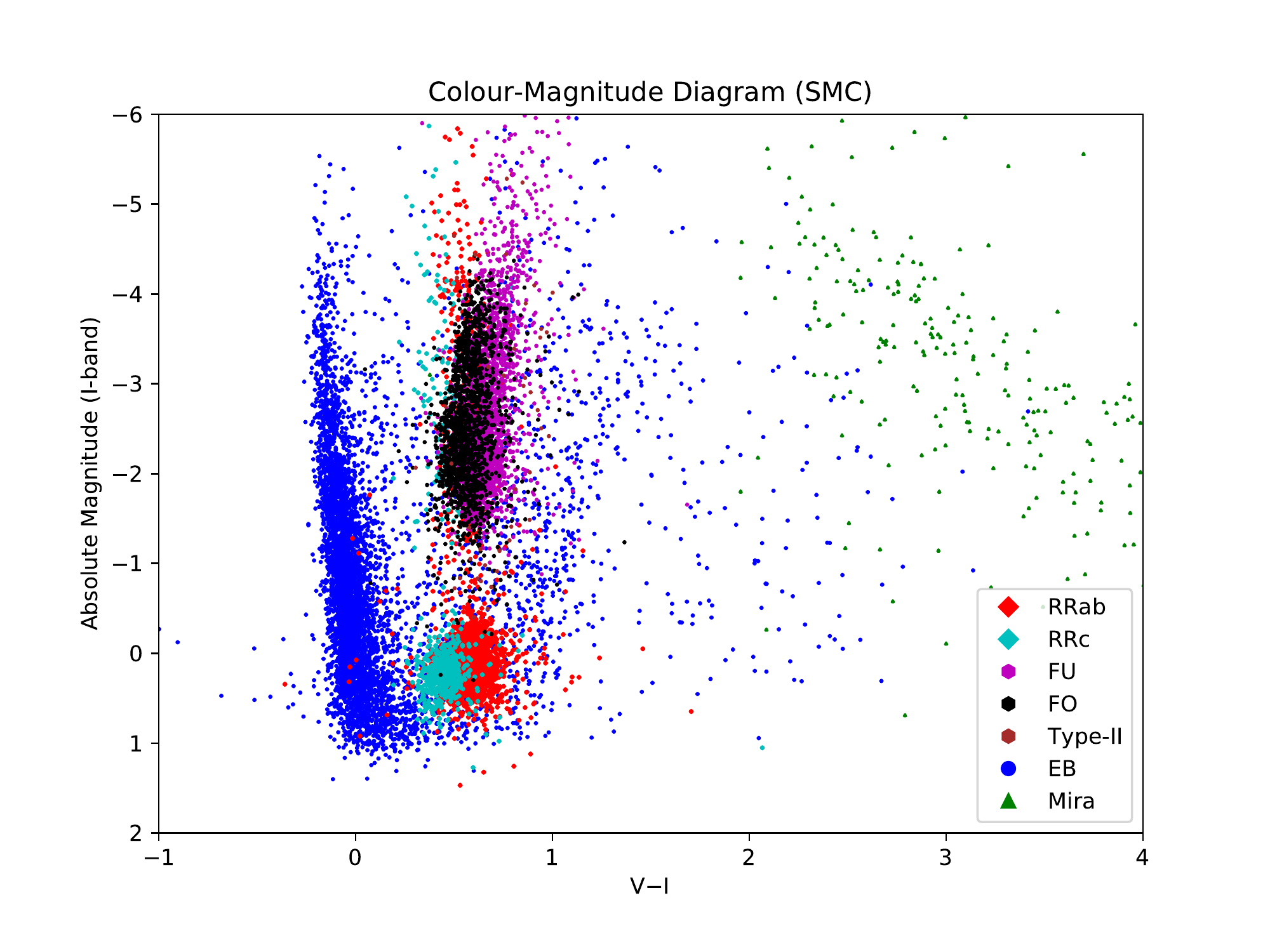}
\caption{Colour-Magnitude diagram for SMC is plotted by applying K-means clustering on six ICs. Different colours represent different variable classes and variables which have similar marker indicator contained in same clusters. }\label{fig:8}
\end{figure*}

\begin{figure*} 
\includegraphics[width=18cm,height=9.5cm]{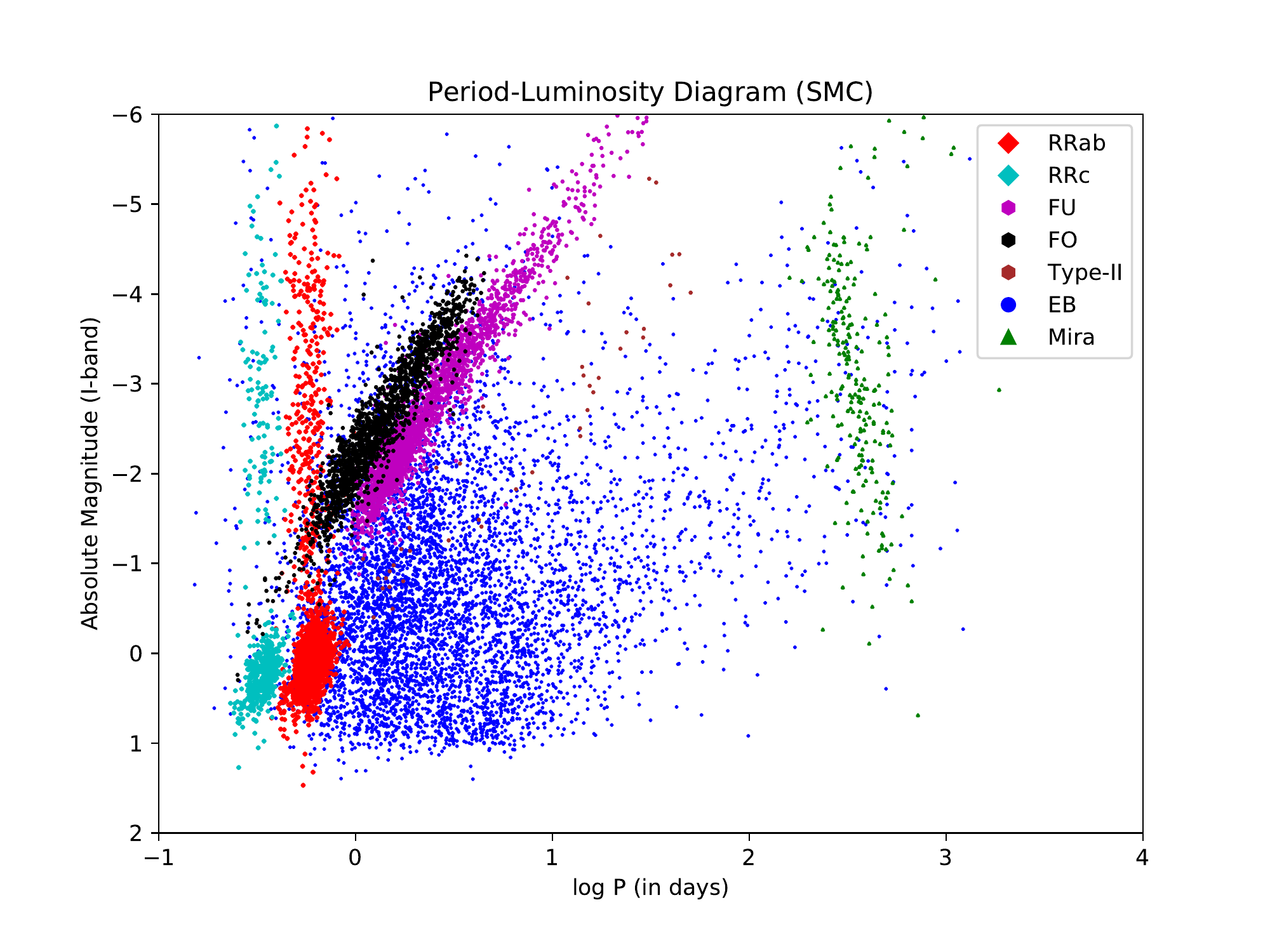}
\caption{Period-Luminosity diagram for SMC is plotted by applying K-means clustering on six ICs. Different colours represent different variable classes and variables which have similar marker indicator contained in same clusters.}\label{fig:9}
\end{figure*}

\begin{figure*} 
\includegraphics[width=18cm,height=10cm]{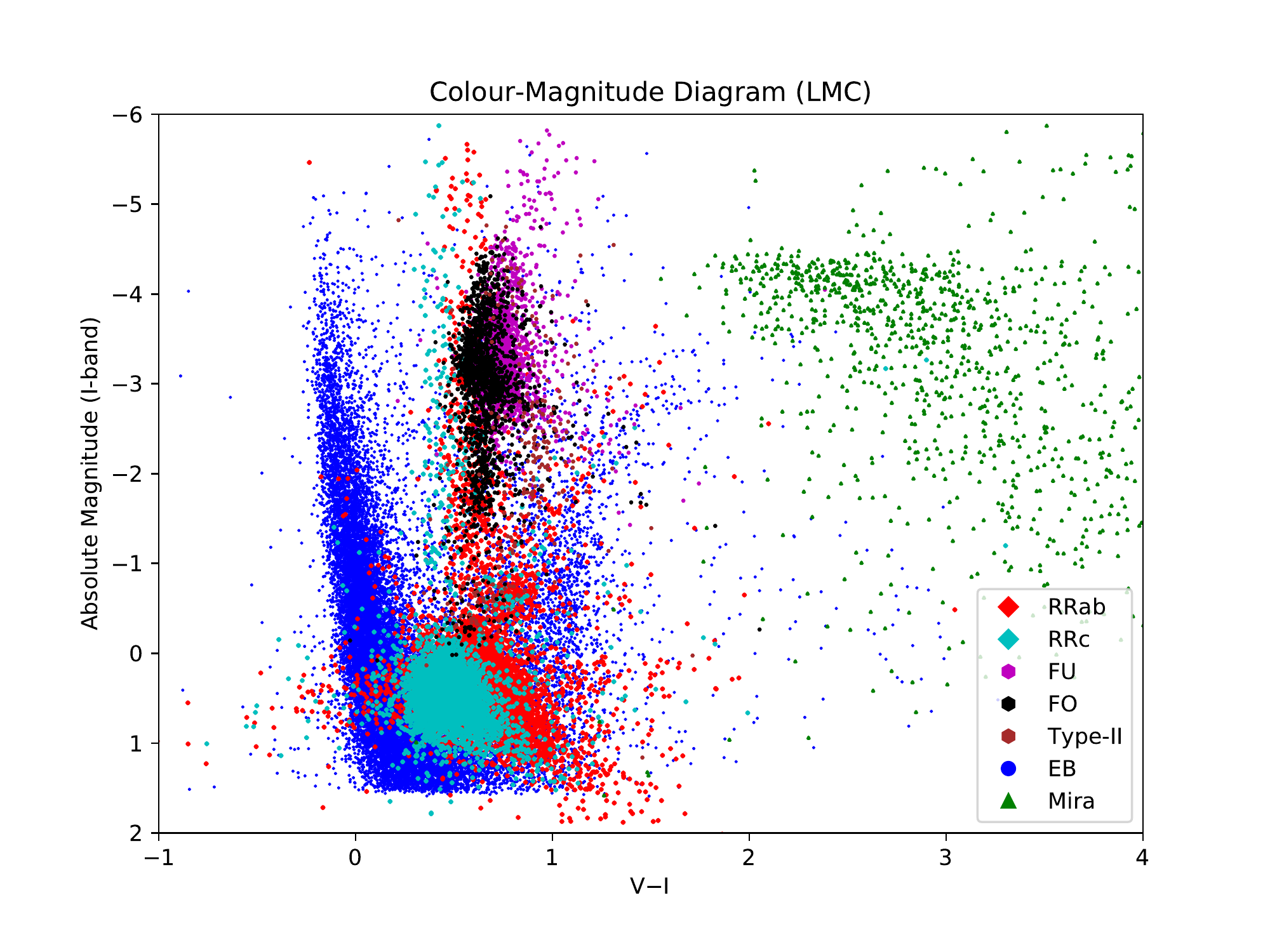}
\caption{Colour-Magnitude diagram for LMC is plotted by applying K-means clustering on five ICs. Different colours represent different variable classes and variables which have similar marker indicator contained in same clusters.}\label{fig:10}
\end{figure*}

\begin{figure*} 
\includegraphics[width=18cm,height=9.5cm]{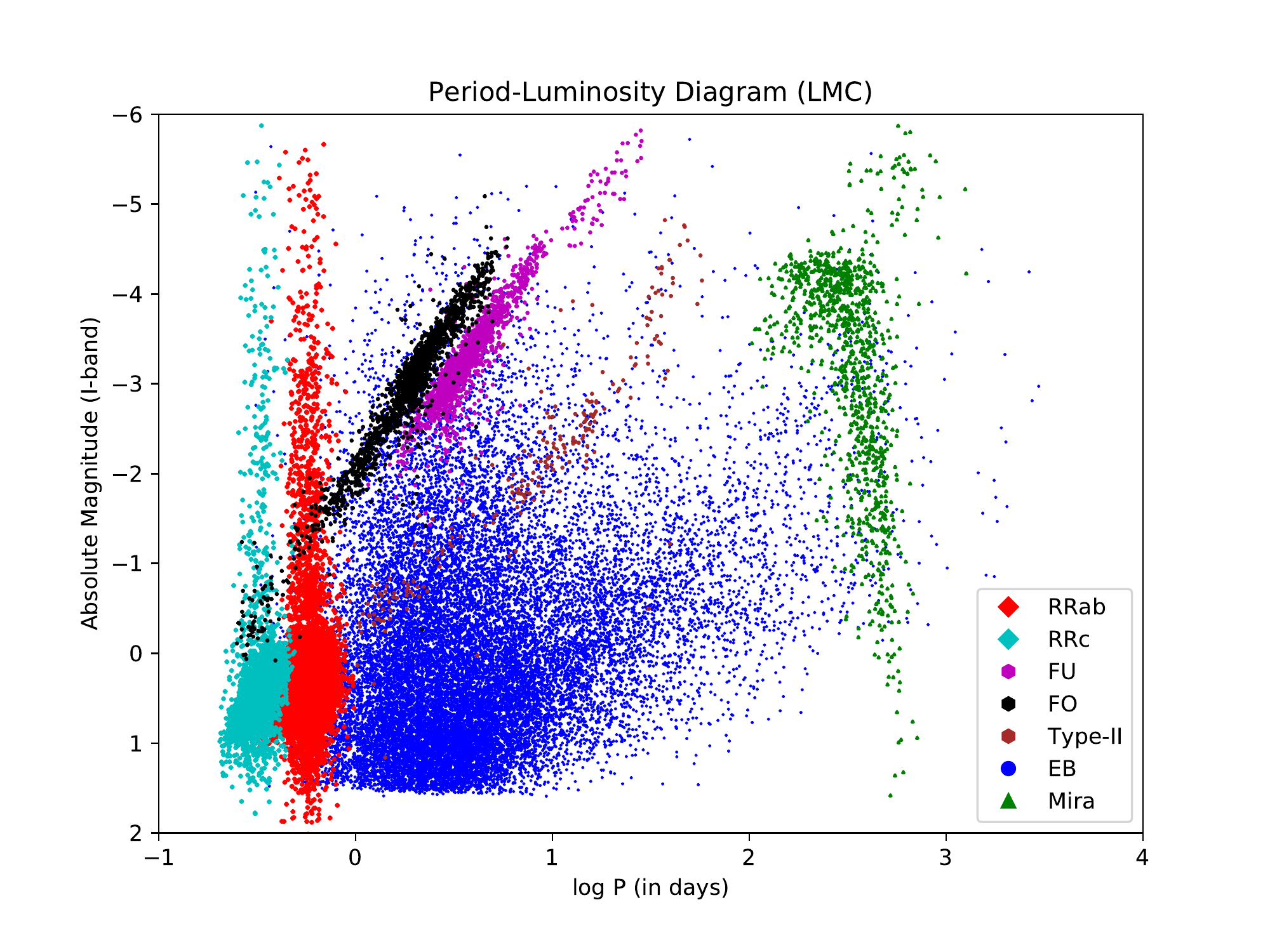}
\caption{Period-Luminosity diagram for LMC is plotted by applying K-means clustering on five ICs. Different colours represent different variable classes and variables which have similar marker indicator contained in same clusters.}\label{fig:11}
\end{figure*}

\begin{figure*} 
\includegraphics[width=18cm,height=10cm]{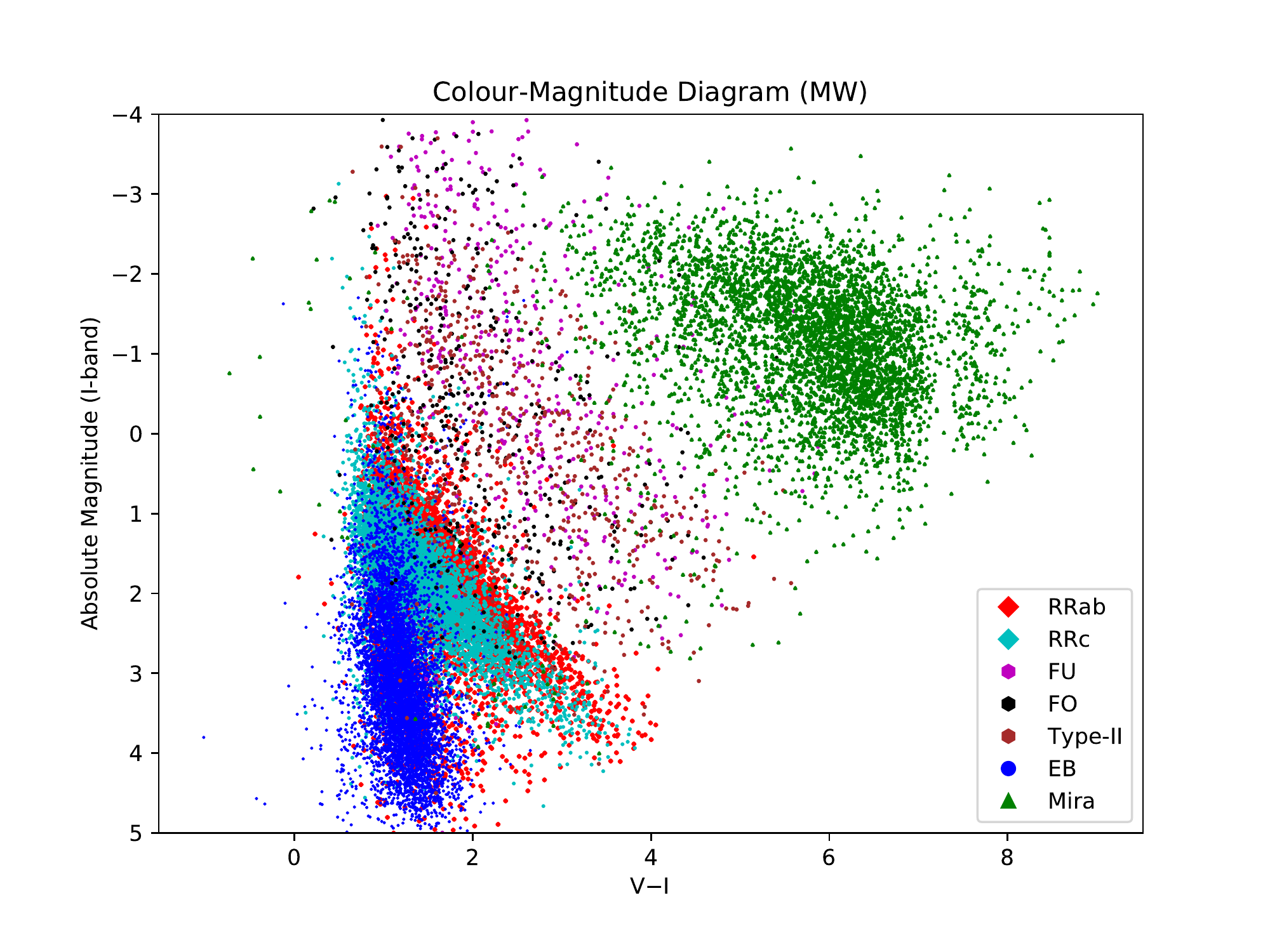}
\caption{Colour-Magnitude diagram for Milky Way is plotted by applying K-means clustering on six ICs. Different colours represent different variable classes and variables which have similar marker indicator contained in same clusters.}\label{fig:12}
\end{figure*}

\begin{figure*} 
\includegraphics[width=18cm,height=9.5cm]{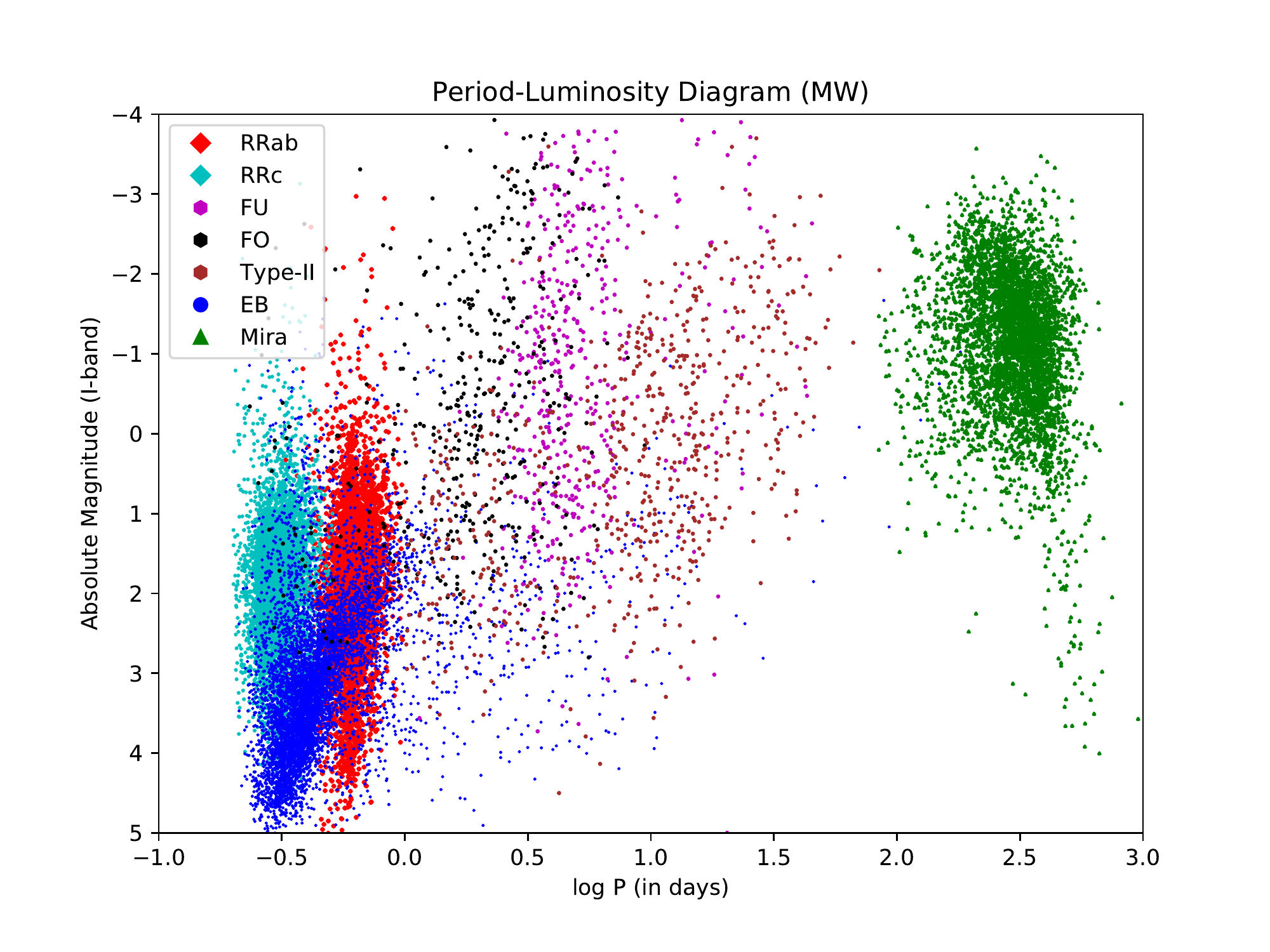}
\caption{Period-Luminosity diagram for Milky Way is plotted by applying K-means clustering on six ICs. Different colours represent different variable classes and variables which have similar marker indicator contained in same clusters.}\label{fig:13}
\end{figure*}

\subsection{K Means clustering with respect to ICs}

In the present work, we have used huge data sets of the light curves (LC) of variable stars in SMC, LMC and MW and have classified the LC by applying CA with respect to the optimum number of IC components. Interestingly we have found four coherent similar groups irrespective of the host galaxies. They are EB (G1), Cepheids (FU, FO, Type-II) (G2), RRab and RRc (G3) and Mira variables (G4).\\ 
The following features have been observed : \\
\\
(i) EB (G1) are mostly main sequence stars and they contain mostly O and B type stars (V-I $\sim$ 0) and their luminosity covers a wide range of spectrum ($M_I \sim$ -1 to -5) i.e this group has a wide range in the mass. Their periods fall in the moderate range (log P $\sim$ 0 to 1). This is consistent with the fact that  in young star clusters the contribution of binary stars is significant (Hu et al. \citeyear{Hu2010}, Wang et al. \citeyear{Wang2020}). Very few stars have evolved from the main sequence branch.\\
\\
(ii) Interestingly all Cepheids (FU, FO and Type-II) fall in one group (G2). All the stars are very bright ($M_I \sim$ -2 to -4) and are comparatively metal rich. The stars have evolved from the main sequence and lie in the Cepheid strip i.e along giant and super giant branches. The Cepheids of FO type have smaller period and lesser magnitudes compared to FU type Cepheids. FU Cepheids are also redder than FO ones. There are slight differences in the magnitudes and colours of Cepheids and EB in SMC, LMC and MW, e.g. SMC contains brighter Cepheids ($M_I \sim$ - 4) than LMC and MW. Also, MW, EB are much redder than SMC and LMC, EB and they have much shorter period (log P $\sim$ - 0.5 to 0) compared to SMC and LMC (log P $\sim$ 0 to 1). This might be due to that they are less massive as seen from their very low luminosity ($M_I \sim$ -2 to 0). The period and luminosity correlation is stronger in SMC and LMC compare to MW.\\
\\
(iii) All RR-Lyrae variables fall in the same group (G3) though RRab have longer period than RRc variables. RR-Lyrae variable stars just turn off from main sequence branch and on the way to giant branch.\\
\\
(iv) It is clear from the color magnitude diagram (CMD) that Mira variables form a separate group (G4) occupying a red giant region of the CMD having brightest magnitudes, reddest colours and longest periods. They are old stars (Baud et al. \citeyear{Baud1981}; Feast \citeyear{Feast2009}) and exceeding main sequence luminosity.  \\
\\
So, it is interesting from the above observation that ICA along with CA can clearly classify the four categories of variable stars with great efficiency and very little amount of overlapping and these four groups are robust i.e independent of their host galaxies. This is the success of the present study. Also the period-luminosity diagram clearly shows the correlation for Type-I Cepheids though the correlation is not very clear for Type-II Cepheids due to small size of the sample. Though the difference between FU and FO Cepheids are not clear in CMD but the difference is clear in period-luminosity diagram as two parallel strips which is only due to the mode of their vibration and not due to chemical composition. This is also true for RRab and RRc variables which form a coherent group.\\
\\
So, we can conclude that ICA along with CA is a very strong and robust tool which can be used for the classification of variable stars and for future prediction on the nature of light curves of variable stars.

\section{Conclusions} \label{sec: Conclusions}
The present work deals with a huge database of OGLE variable stars light curve data (RR Lyrae, Cepheids, EB, Mira etc.). It consists of Heliocentric Julian Days (HJD), observed I-band magnitudes. Using spline interpolation, we obtain 1000 magnitudes for each light curve between phase 0 to 1 with step length 0.001. The input to PCA and ICA is the matrix form of normalized magnitude data set obtained using equation (\ref{eqn:5}).\\
\\
The following features have been observed :\\
\\
(i) PCA and ICA have been applied separately to matrices of order 2428 $\times$ 1000, 2739 $\times$ 1000 and 1184 $\times$ 1000 corresponding to light curves for FU Cepheids for LMC, SMC and MW respectively and we observed that the overall performance of ICA is better than PCA and FD (Deb $\&$ Singh \citeyear{Deb2009}) for finding out resonances in FU cepheids in LMC and SMC but for MW the performance of ICA and PCA is similar. For FO Cepheids, the input matrices are 1766 $\times$ 1000, 1783 $\times$ 1000 and 542 $\times$ 1000 towards LMC, SMC and MW respectively and we find out that the performance of PCA and ICA is similar for LMC and SMC but for MW the overall performance of ICA is better for bringing out the structural changes in the shape of FO Cepheids with respect to  their periods.\\
\\
(ii) We demonstrate the capabilities of PCA and ICA in classifying different variable stars for LMC, SMC and MW separately. For SMC the input matrix for PCA and ICA is of order 18889 $\times$ 1000. From Fig. \ref{fig:5}, it is clear that IC3 outperforms PC1 not only in the separation of FU and FO Cepheids but also separation between eclipsing binary to other variables except for Mira variables due to some outlier present in the data. Similarly, IC2 performs much better than PC1 in the case of LMC where the input matrix is of order 71025 $\times$ 1000. There is no such difference in the performance of ICA and PCA in the case of MW where the input matrix is of order 54017 $\times$ 1000.\\
\\
(iii) Finally, applying K-Means clustering on ICs we have plotted the period-luminosity diagram and the colour-magnitude diagram separately for LMC, SMC and MW galaxies. Interestingly ICA along with CA introduce four categories of variable stars namely EB (G1), Cepheids (G2), RR Lyrae (G3) and Mira (G4), independent of host galaxies as seen from Fig. (\ref{fig:8}) - (\ref{fig:13}).\\
\\
Therefore we conclude by saying that ICA is a dimension reduction technique which has the capability of identifying possible resonances for large light curve data sets of FU and FO Cepheid and performs better than PCA. Also, ICA outperforms PCA and FD method (Deb $\&$ Singh \citeyear{Deb2009}) in classifying big data sets, which will be more suitable when dealing with larger databases with a more diverse set of variable stars in future. Also, ICA along with CA is a robust classifier for variable stars helping for future predictions on the nature of variable stars.

\section*{Acknowledgements}

Author S.P acknowledges INSPIRE SRF Vide sanction Order No. DST/INSPIRE Fellowship/2017/IF170368 under the Department of Science and Technology (DST) INSPIRE program, Government of India. The authors are grateful to Mr. Prasenjit Banerjee for providing valuable support.

\section*{Data availability}

The data underlying this article are available in OGLE On-line Data Directory, at \url{https://ogle.astrouw.edu.pl/}.

\bibliographystyle{mnras}
\bibliography{example} 
\bsp	
\label{lastpage}
\end{document}